\documentstyle[12pt,aaspp]{article}

\def\etal{et al. }
\def\simlt{\lower.5ex\hbox{$\; \buildrel < \over \sim \;$}}
\def\simgt{\lower.5ex\hbox{$\; \buildrel > \over \sim \;$}}
\def\COBE {{\it COBE}}
\def\IRAS {{\it IRAS}}

\begin{document}

\title{FAR-INFRARED SPECTRAL OBSERVATIONS OF THE GALAXY
BY COBE\footnote{The National Aeronautics and Space Administration/
Goddard Space Flight Center (NASA/GSFC) is responsible for the
design, development, and operation of the Cosmic Background
Explorer (\COBE). Scientific guidance is provided by the \COBE\
Science Working Group. GSFC is also responsible for the development
of the analysis software and for the production of the mission
data sets.}}\par

\author{W. T. Reach\altaffilmark{1}, 
        E. Dwek\altaffilmark{2}, 
        D. J. Fixsen\altaffilmark{3}, 
        T. Hewagama\altaffilmark{4}, 
        J. C. Mather\altaffilmark{2}, 
        R. A. Shafer\altaffilmark{2},
        A. J. Banday\altaffilmark{1}, 
        C. L. Bennett\altaffilmark{2}, 
        E. S. Cheng\altaffilmark{2}, 
        R. E. Eplee, Jr.\altaffilmark{5},
        D. Leisawitz\altaffilmark{6}, 
        P. M. Lubin\altaffilmark{7},
        S. M. Read\altaffilmark{4}, 
        L. P. Rosen\altaffilmark{4}, 
        F. G. D. Shuman\altaffilmark{4},
        G. F. Smoot\altaffilmark{8},
        T. J. Sodroski\altaffilmark{3},
\&      E. L. Wright\altaffilmark{9}
}
\altaffiltext{1}{Universities Space Research Association,
        NASA Goddard Space Flight Center, Code 685, Greenbelt, MD 20771}
\altaffiltext{2}{NASA Goddard Space Flight Center, Code 685, Greenbelt, MD 20771}
\altaffiltext{3}{Applied Research Corporation, Code 685.3, Greenbelt, MD 20771}
\altaffiltext{4}{Hughes-STX, Code 685.9, Greenbelt, MD 20771}
\altaffiltext{5}{General Sciences Corporation, Code 685.3, Greenbelt, MD 20771}
\altaffiltext{6}{NASA Goddard Space Flight Center, Code 631.0, Greenbelt, MD 20771}
\altaffiltext{7}{UCSB Physics Department, Santa Barbara, CS 93106}
\altaffiltext{8}{LBL \& SSL, Bldg 50-351, University of California, Berkeley, CA 94720}
\altaffiltext{9}{UCLA Astronomy Department, Los Angeles, CA 90024-1562}


\begin{abstract}
We derive Galactic continuum spectra from 5--96 cm$^{-1}$ from
COBE/FIRAS observations. The spectra are dominated by warm dust
emission, which may be fit with a single temperature 
in the range 16--21 K (for $\nu^2$ emissivity)
along each line of sight.
Dust heated by the attenuated radiation field in molecular clouds
gives rise to intermediate temperature (10--14 K) emission in
the inner Galaxy only.
A widespread, very cold component (4--7 K) with
optical depth that is spatially correlated with the warm component
is also detected. The cold component is unlikely to be due to 
very cold dust shielded from starlight, because it is present
at high latitude. We consider hypotheses that the cold component 
is due to enhanced submillimeter emissivity of the dust that gives 
rise to the warm component, or that it may be due to very small,
large, or fractal particles.
Lack of substantial power above the emission from warm dust places
strong constraints on the amount of cold gas in the Galaxy.
The microwave sky brightness due to interstellar dust is dominated
by the cold component, and its angular variation
could limit our ability to discern primordial fluctuations in the 
cosmic microwave background radiation.
\end{abstract}

\keywords{Infrared: Interstellar: Continuum --- ISM: Dust, Extinction --- 
Radio Continuum: Interstellar}

\section{Introduction}

Absolute spectrophotometry on large angular scales in the far-infrared 
($\lambda \simgt 100\mu$m) and submillimeter 
($\lambda \simlt 1$mm) bands is not possible using 
ground-based instruments, because of the large, time-varying 
attenuation by the Earth's atmosphere. 
Even using balloon-based instruments, observations require 
frequent chopping to `blank' regions, so that the accurate
photometry is available only on scales comparable to the
reference beam separation. 
Portions of the galactic plane were surveyed by
balloon instruments in wavebands between 150 and 300$\mu$m,
revealing the bright ridge of dust emission from H~II regions
and molecular and atomic clouds centered on
the galactic plane (cf. Hauser \etal 1984 and references therein).
The Infrared Astronomical Satellite (IRAS) made
the first all-sky survey at 100$\mu$m wavelength 
(Neugebauer \etal 1984). The 100$\mu$m emission at high
galactic latitude is strongly correlated to the distribution of
atomic gas as traced by the 21-cm line of H~I (Boulanger
\& Perault 1987), and can be explained by emission from dust grains,
mixed with the interstellar gas and heated by the interstellar
radiation field (Draine \& Lee 1984).
Interstellar grains of various sizes contribute different parts of
the infrared emission. In one recent model, the emission is broken
into three components, due to `big' grains, `very small' grains, 
and polycyclic aromatic hydrocarbons (D\'esert, Boulanger, \&
Puget 1990). The far-infrared emission is dominated by `big' grains,
which also dominate the mass and total luminosity of interstellar
dust.

In this paper we describe all-sky observations by the 
Far Infrared Absolute Spectrophotometer (FIRAS)
aboard the Cosmic Background Explorer (\COBE),
which provide the first complete coverage of the wavelength
range from 4.5mm through 104$\mu$m. Preliminary results on the
Galactic continuum and spectral lines in this wavelength range 
were presented by Wright \etal (1991), and the angular variation
of the brightnesses of the spectral lines was presented
by Bennett \etal (1994).
Here we focus on the
galactic continuum spectrum and its variation both in the galactic
plane and at high galactic latitude.

\section{Observations}

\subsection{FIRAS Description}

The FIRAS is one of the three Cosmic Background Explorer 
(\COBE) instruments. During its observing 
lifetime (November 1989 to September 1990), 
95\% of the sky was covered by scans at a solar elongation of 91--94$^\circ$.
The effective beam has a FWHM of  $7^\circ$
and drops sharply to -30 db by $10^\circ$ off-axis, as confirmed
by observations of the Moon. The spectrometer is a
Michelson interferometer with bolometric detectors sensitive to
1--96 cm$^{-1}$ radiation. 
(Throughout this paper, we use the
cm$^{-1}$ scale for frequencies, $\nu$. To convert to GHz, multiply 
by 30; to convert to wavelength in $\mu$m, use $10^4/\nu$.) 
The incoming photons are separated into high- and low-frequency 
channels by a dichroic filter. The spectrometer was operated in
several modes by varying the stroke length and scan rate. For
this paper the data from the low and high frequency detectors
were combined, so that the spectra cover frequencies 2.2--96 cm$^{-1}$ 
(or wavelengths 4.5mm--104$\mu$m), with a spectral resolution of
0.57 cm$^{-1}$ over the entire range.
(See Fixsen \etal 1994 for technical details). 
The observations were absolutely calibrated
by placing a beam-filling blackbody in the aperture during
calibration periods (once per month during most of the mission).
Relative calibration was maintained by always measuring the difference
between the sky signal and that of an internal calibrator.

\subsection{Data Preparation}

The monopole and dipole components of the cosmic background 
radiation were removed from each spectrum
by subtracting a blackbody with a temperature 
$T_{CBR}=T_0+T_d\cos\Theta$, where $T_0=2.726$~K, $T_d=3.343$~mK
(Mather \etal 1993), 
and $\Theta$ is the angle between the line of sight and the
orientation of the cosmic dipole $(l=264.4^\circ,b=48.4^\circ)$
(Kogut \etal 1993). 
The resulting spectra are dominated by the continuum emission
of galactic dust, spectral lines of galactic gas, 
the (as-yet undetected) cosmic infrared background,
and contributions from the zodiacal light at high frequencies
and galactic synchrotron and free-free emission at very low frequencies.

The zodiacal light was modeled using observations by the \COBE\ Diffuse 
Infrared Background Experiment (DIRBE).
The DIRBE observations at 5 wavelengths between 4.9 and 100$\mu$m were
used to fit a parameterized model of the distribution and temperature
of interplanetary dust. 
The solar elongation of the DIRBE line of sight varied from 
$64^\circ$ to $124^\circ$ during each spin about the \COBE\ rotation
axis, introducing a strong time dependence to the sky brightness.
The parameterized model was optimized to match the temporal
variation of the sky brightness; therefore, no assumptions were
made about the time-invariant sky brightness outside the Solar System.
The parameterized zodiacal light model was evaluated for each FIRAS 
line of sight, and for a frequency of 70 cm$^{-1}$, to produce a 
spatial template, $Z(l,b)$. 
This spatial template was then used to determine the zodiacal light 
spectrum, by assuming the brightness can be decomposed into cosmic,
Galactic, and zodiacal components:
\begin{equation}
I_\nu(l,b) = B_\nu (T_{cbr}) + g_\nu G(l,b) + z_\nu Z(l,b),
\label{tempeq}
\end{equation}
where the cosmic microwave background is assumed to be the only
isotropic background. The decomposition of the FIRAS spectrum for
the region $(l=20\pm 5^\circ, b=20\pm 5^\circ)$ 
is shown in Figure~\ref{figcgz}.
This technique is similar to that used by Wright \etal (1991) to 
determine the spectrum of the Galaxy, $g_\nu$; for this paper, we 
excluded the galactic plane ($|b|<30^\circ$) and added the zodiacal term. 
The resulting zodiacal light spectrum is adequately represented by 
a power-law, $z(\nu) \propto \nu^3$, over the FIRAS bandpass. 
We removed the zodiacal light from each FIRAS spectrum by subtracting
$(\nu/70)^3 Z(l,b)$. 
If the zodiacal spatial template is accurate to better than 10\%, then the
zodiacal light subtraction introduces negligible uncertainty
except at the very highest frequencies and at the galactic pole.
At high galactic latitude ($|b|>60^\circ$), the zodiacal light 
contributes 60\% of the total emission at 100 cm$^{-1}$, 
25\% at 70 cm$^{-1}$, and 10\% at 42 cm$^{-1}$.

The contributions of Galactic synchrotron and free-free emission 
to the FIRAS frequencies were constrained using observations 
by the \COBE\ Differential Microwave Radiometer (DMR). 
The DMR observations at three frequencies, 31, 53, \& 90 GHz
(1.0, 1.8, \& 3.0 cm$^{-1}$), 
were extrapolated to higher frequencies assuming the spectrum
follows a power-law, $S_\nu\propto \nu^{-\beta}$.
For synchrotron emission, the spectrum follows $\beta\simeq 0.75$ at
lower frequency (cf. de Bernardis, Masi, \& Vittorio 1991), 
and then the spectral index steepens to
$\beta\sim 1$ above 10 GHz (Banday \& Wolfendale 1991).
For free-free emission, the spectral index is $\beta=0.1$.
In order to place an upper limit to the free-free
and synchrotron emission at higher frequencies, we used the
shallower $\beta=0.1$ to scale the total brightness at 53 GHz.
This is very conservative, because a substantial fraction of
the 53 GHz emission is due to synchrotron and dust emission.
A cosecant of Galactic latitude was fitted to the 53 GHz map, 
and an isotropic component with a brightness equal to the 
cosecant slope (12$\mu$K) was added to the map to replace 
the isotropic component (cf. Bennett \etal 1992).
The combined synchrotron and free-free emission account for
$< 10$\% of the emission at frequencies higher than 5 cm$^{-1}$.
In what follows, we excluded frequencies below 5 cm$^{-1}$ 
in order to avoid any possibility of contamination by free-free
emission.

There were 9 spectral lines detected by FIRAS 
in the average spectrum of the Galaxy 
(Wright \etal 1991, Bennett \etal 1994),
and several of these are evident in each spectrum at low galactic
latitude. These lines were avoided for the purposes of this
paper by eliminating the appropriate frequencies.
All of the lines are unresolved, so only two frequency bins 
contain each line emission; however, some ringing in
the apodization function was evident for the very bright lines.
We removed spectral frequencies within 1.5 cm$^{-1}$ of the 
bright C~II (63.395 cm$^{-1}$) and N~II (48.72 and 82.04 cm$^{-1}$)
lines, and within 1 cm$^{-1}$ of the C~I (16.419 and 27.00
cm$^{-1}$) and CO (7.681, 11.531, 15.375, and 19.237 cm$^{-1}$) 
lines. 

Coaddition was necessary in order to accurately measure the
galactic spectrum at high latitudes.
The spectra were combined into 120 longitude bins in the 
galactic plane and 26 high-latitude regions.
The bins were chosen to be 
larger at higher latitude, where the signal is relatively weak.
For the galactic plane spectra, the bins are evenly spaced
bins at 3$^\circ$ longitude intervals and include data with 
$|b| < 3^\circ$. The high latitude
bins consist of the north and south galactic polar caps,
with $|b|>60^\circ$, high-latitude zones with
$60^\circ>|b|>30^\circ$ in 90$^\circ$-wide longitude bins,
and intermediate-latitude zones with $30^\circ>|b|>10^\circ$
in $45^\circ$-wide longitude bins. 
Two sample spectra, toward ($l=45^\circ,b=0^\circ$) and 
$(180^\circ>l>90^\circ, -30^\circ>b>-60^\circ$)
are shown in Figures~\ref{fitfig}(a) and ~\ref{fitfigh}(a).
The error bars shown here and used in the spectral analysis are
the calibration uncertainties in the mean brightness over each 
spatial region,
and they do not reflect the intrinsic variation of the brightness 
within the region.

\subsection{Spectral Analysis}

The spectrum for each region was fitted by one or more
modified blackbodies of the form
\begin{equation}
I_\nu = \tau_0 \epsilon_\nu B_\nu(T),
\label{modbb}
\end{equation}
where $\tau_0$ is the optical depth at $\nu_0=30$~cm$^{-1}$
($\lambda_0=333\mu$m), and $\epsilon_\nu$ is the emissivity
normalized to unity at $\nu_0$.
The simplest fit for each spectrum was a single
component with a power-law emissivity, 
$\epsilon_\nu = (\nu/\nu_0)^\alpha$. 
In the low-frequency limit, the Kramers-Kronig theorem can be used to
show that $\alpha$ is an even positive integer (Wright 1993),
although it is not yet known below which frequency this limit applies.
The simplest fit we considered was for a $\nu^2$ emissivity,
so that there are only 2 free parameters ($T$ and $\tau$).
As examples, the residual spectra
following two such fits are shown in Figures~\ref{fitfig}(b) and 
~\ref{fitfigh}(b). The residuals clearly indicate the presence 
of excess emission at low frequencies in nearly all of the spectra, 
indicating a very cold emission component that is widespread. 
We repeated the spectral fits allowing $\alpha$ to vary, which
substantially improved the $\chi^2$ of most fits.
Examples of the residuals that resulted in this case are
shown in Figures~\ref{fitfig}(c) and ~\ref{fitfigh}(c).

In the inner Galaxy, where the signal to noise is high,
poor fits were obtained with a power-law emissivity.
The galactic center spectrum
at frequencies below 50 cm$^{-1}$ is best fit by 
$\alpha=1.98\pm 0.01$, and the other spectra require lower $\alpha$.
There are good reasons to expect that the value of
$\alpha$ derived from fits using single temperatures
is lower than the true value:
the distribution of temperatures along the line of sight should
broaden the spectrum, and lower our derived $\alpha$.
The highest observed value of $\alpha$ is most likely
to approach the true value, and we will therefore adopt
$\alpha=2$ at low frequencies. At higher frequencies,
the emissivity increases less steeply than $\nu^2$. 
A broken power-law model that can match the emissivity rolloff
at high frequencies has $\alpha=1.5$
for frequencies above 70 cm$^{-1}$; however, the broken
power-law introduces an unobserved cusp so that 
$\chi^2$ is large for these fits.
A smooth model for the emissivity variation that can match the 
high-frequency rolloff and improve the $\chi^2$ of the fits is
\begin{equation}
\epsilon_\nu = { {(\nu/\nu_0)^2}\over{ [1 + (\nu/\nu_1)^\beta]^\frac{1}{\beta} } },
\label{emiss}
\end{equation}
with $\nu_1\sim 50$ cm$^{-1}$ and $\beta\sim 6$. 
These values of $\nu_1$ and the exponents
in equation~\ref{emiss} were chosen by experiment, and are not the
result of a formal fit; they improve $\chi^2$ for the $l=15^\circ$
fit by a factor of 2.
The high-frequency rolloff is needed only in the inner galaxy, but
is not excluded in the high-latitude spectra, where the signal-to-noise
is low.
At high latitude, a simple power-law
($\nu^2$) is adequate. It cannot be determined with the present data
whether the high-frequency rolloff in the emissivity is due to
different grain properties in the inner galaxy, or
the distribution of grain temperatures along the line of sight.
In what follows, we use equation~\ref{emiss} for the galactic plane
and a simple $\nu^2$ for the high-latitude spectra.

\section{Results}

\subsection{Significance of the spectral models}

The FIRAS spectra of galactic emission can be characterized by 
three components,
which we call `warm' ($16 < T_1 < 21$~K), `cold' ($4 < T_2 < 7$~K),
and `intermediate' ($10 < T_3 < 13$~K). There is no overlap
between the temperatures of the three components, and the ranges
are those of the fit results.
Each spectrum was fitted by a single-component, $\alpha=2$ model 
($T$, $\tau$ varying); a single-component, free $\alpha$ ($T$,
$\tau$, $\alpha$ varying); and a two-component, $\alpha=2$ model
($T_1$, $\tau_1$, $T_2$, $\tau_2$ varying), with the goal of
finding the simplest model that could fit the most spectra.
The significance of a given model was judged by its improvement
in the reduced $\chi^2$, according to the F-test.
The probability that the improvement in $\chi^2$ due to including
an extra parameter was {\it not} due to chance, $P(>F)$, was
calculated.
In every case, the single-component, free-$\alpha$ model and the
two-component models were significantly better than the single-component,
$\alpha=2$ model, and in most cases, the two-component 
model was significantly better than either single-component model.
In Table~\ref{planetab}, the results of the two-component model 
for the galactic plane spectra are listed, 
together with the probability, $P(>F_{ac})$, that the 
two-component model is significantly better than the single-component,
$\alpha=2$ model, 
and the probability, $P(>F_{bc})$, that the two-component model is
significantly better than the single-component, free $\alpha$ model.
For some spectra, it was not possible to distinguish between
the single-component, free $\alpha$ and two-component models at the 
90\% confidence level.
In the inner Galaxy, both the one-  and two-component fits are poor 
because of the intermediate-temperature emission. A three-component fit
improves $\chi^2$ in the inner galaxy, but cannot be fitted to the
outer galaxy or high-latitude spectra because the third component
cannot be separated from the other two. We therefore
consider the two-component model for the analysis of 
the galactic plane spectra.

The results of some of the high-latitude fits are shown in 
Tables~\ref{highlattab}a, b, and c for the various spectral models. 
The excess submillimeter emission above a $\nu^2$ modified blackbody
is present with $> 95$\% confidence in all spectra.
The distinction between the single-component, free $\alpha$ model
and the two-component model
was not possible for many high-latitude spectra. 
The results of the free-$\alpha$ fit indicate statistically significant 
variations in $\alpha$, so that the emissivity law varies 
substantially from place to place.
Based on the fact that the
two-component model was more successful in the galactic plane,
it is likely that the modest amount of cold emission,
even at $|b|>60^\circ$, is also due to a second component.

\subsection{Effect of a cosmic infrared background}

The cosmic infrared background radiation (CIBR) was assumed to be
small in this analysis, although the observational upper limits to
its brightness still allow it to contribute a significant fraction
of the sky brightness. Models of the integrated light of galaxies
(Franceschini \etal 1994)
predict CIBR brightnesses that are 5--25\% of the 10-100 cm$^{-1}$
sky brightness (after zodiacal light subtraction) at the galactic poles.
In order to determine whether the presence of a CIBR would have a 
significant effect on our conclusions, we subtracted the brightest
CIBR from Franceschini \etal (1994) from the FIRAS spectra and
repeated the spectral fits.
The optical depths decreased, as expected because the sky is fainter,
but the temperatures of the warm and very cold components are not
significantly affected. We therefore feel confident in neglecting
the CIBR, noting only that the dust optical depth at the highest latitudes
is overestimated.

\subsection{Effect of a temperature distribution}

Two effects will produce a range of temperatures in each spectrum.
First, each particle size has its own temperature, determined 
by the balance between its far-infrared and visible absorption efficiencies. 
Second, different regions of the Galaxy have different radiation
fields, which heat the dust to different temperatures.
The issue of disentangling these effects is beyond our present scope.
The radial variation of the interstellar radiation field can best be
determined by comparison to spectral-line
surveys of CO and H~I, for which the distances can be determined from
Galactic rotation.
In order to gain some insight into the effect of a temperature distribution,
we examined the following spectral model:
\begin{equation}
I_\nu = \tau_0 \epsilon_\nu \frac{1}{\Delta T}
        \int_{T_0-\Delta T}^{T_0+\Delta T} B_\nu(T) dT,
\label{eqdis}
\end{equation}
which assumes equal probabilities for grains at each temperature
in the range $T_0\pm \Delta T$. 
There is a strong covariance between $T$ and
$\Delta T$, such that opening the width of the temperature distribution
leads to lower mean temperatures. 
We have fitted eq.~\ref{eqdis} to representative FIRAS spectra for the
inner Galaxy, outer Galaxy, and the high-latitude sky. 
For the inner Galaxy, the optimum width to the temperature
distribution was $\Delta T=0$. 
The fact that eq.~\ref{eqdis}
does not improve the fits, despite adding an extra degree of freedom,
leads us to conclude that either the temperature distribution
is not important, or it has a substantially different functional form
than we considered.
For the outer Galaxy, eq.~\ref{eqdis} produced fits of comparable
quality ($\chi^2$) for a wide range of $\Delta T$, as long as the 
maximum temperature, $T_0+\Delta T\simlt 19$~K. Thus it is possible
that comparable column densities of dust exist at a wide range of
temperatures in the outer Galaxy. The power emitted depends 
approximately on the 5--6$^{th}$ power of temperature, so 
negligible amounts of power are emitted by dust much cooler than
that at the highest temperatures. 
At high latitudes, eq.~{eqdis} was no better than a single temperature.
It is worth noting that changing $\Delta T$ has only a small effect
on the shape of the spectrum at low frequencies. Thus using
a single temperature, rather than a continuous distribution,
for the warm dust emission has no impact on 
the presence of the `very cold' component in the spectra.

The small width of the temperature distribution for the inner
Galaxy has direct implications for the
variation of temperature with galactocentric radius and particle size.
An upper limit to the width of the temperature distribution for the
line of sight in the Galactic plane toward $l=45^\circ$
is $\Delta T < 2$~K at the 95\% confidence level. 
The expected $\Delta T$ due to Galactic radial variations may be 
estimated by assuming the volume emissivity scales exponentially
with scale length $R_\epsilon$, and 
and the radiation field has scale length $R_\star$; 
then $\Delta T/T \simeq R_\epsilon/6R_\star$.
If $R_\epsilon =3.1$~kpc, $R_\star =3.5$~kpc, and $T\simeq 16$~K, 
then $\Delta T\simeq 2$~K, comparable to our upper limit to $\Delta T$. 
The expected $\Delta T$ due to the range of particle sizes may be 
estimated as the rms dispersion in $T$ weighted by the emissivity
per particle size; we find $\Delta T\simeq 1$~K from this effect.
Thus the bulk of the far-infrared emission is fit well 
by a narrow temperature range, and the expected temperature
range due to the Galactic radial gradient and the size distribution
is comparable to the upper limit derived from the data.
This justifies the assumption of a single temperature along the line
of sight as a first approximation.

\subsection{Distribution of temperature and optical depth}

The longitude profiles of the optical depths and temperatures 
of the warm and cold emission are shown in Figure~\ref{ttaufig}.
It is evident that the temperature of the warm component varies
strongly with galactic longitude. This variation is expected because
lines of sight passing closer to the galactic center sample dust 
subjected to a higher radiation field (Mathis, Mezger, \& Panagia 1983).
The temperature of the cold component does not show the same
longitude variation; if anything, the temperature of the cold
component may be lower in the inner Galaxy.

The high-latitude variations of the temperature and optical depth 
of the `warm' and `cold' components are shown schematically in 
Figure~\ref{mapparfig}.
In the $10^\circ <|b|<30^\circ$ zone, the temperature of the warm component 
generally follows that in the galactic plane, with clear exceptions
being the relatively hot Orion starforming region at
$(l=210^\circ, b=-20^\circ)$ and the relatively cold Taurus dark
cloud region at $(173^\circ,-14^\circ)$. Note that the optical
depth is high in both the Orion and Taurus regions, demonstrating
that the temperature and optical depth vary independently and
can be separately determined.
The Orion region contains a large OB association, and has
a radiation field enhanced by a factor of $\simgt 2$ 
with respect to the general interstellar radiation field,
when averaged over scales comparable to the FIRAS beam 
(Wall \etal 1994). The Taurus region has relatively few OB stars,
and its infrared emission is dominated by relatively dense,
cold molecular clouds.
At higher latitudes, the temperature approaches the polar value
of around 18~K, with both poles comparable. 

The optical depths of the warm and cold components are
well correlated. In Figure~\ref{tautaufig}, they are plotted
against each other. The rank correlation coefficient between $\tau_1$
and $\tau_2$ is 0.96 for all spectra, and it is 0.95 for the
high-latitude spectra only. The correlation is not quite linear,
with the average $\tau_2/\tau_1=7.1\pm 0.4$ for high latitudes
(low optical depths) and $5.2\pm 0.2$ for low latitudes
(high optical depths). It is remarkable that the cold and warm
components are so well correlated, and that nearly the same
ratio of optical depths is obtained for all lines of sight.
However, it is also significant that the correlation is not
perfect, so there is not a single, universal type of dust.

\subsection{Comparison to Previous Observations}

The distinction between the warm and very cold components was 
made possible by the continuous spectral coverage of FIRAS,
extending throughout the far-infrared and submillimeter bands. 
Observations in the 100--350$\mu$m range
are consistent with dust at a single temperature---the warm dust.
The warm dust temperatures we measure in the galactic plane are similar 
to those
obtained from balloon-based surveys (cf. Hauser \etal 1984), and
the high-latitude dust temperature we measure is similar to that
($16.2^{+2.3}_{-1.8}$ K) obtained from broadband 
(134, 154, \& 186$\mu$m) observations by a Berkeley/Nagoya
rocket experiment (Kawada \etal 1994).
The very cold emission has been detected before, and ascribed either to
a submillimeter excess or to an emissivity index $\alpha<2$.
Combining broad-band observations at 5.6, 8.7, 15.8, and 22.5 cm$^{-1}$
(1800, 1100, 630, and 440 $\mu$m) by a MIT balloon-borne instrument
(Page, Cheng, \& Meyer 1990)
with observations at 33, 40, and 67 cm$^{-1}$ (300, 250, and 150 $\mu$m)
by a Goddard balloon-borne instrument (Hauser \etal 1984)
and observations at 100 and 167 cm$^{-1}$ (100 and 60 $\mu$m) by IRAS,
the spectrum of the galactic plane at $l=42^\circ$ was found to
have a submillimeter excess (Page \etal 1990). 
The excess could be characterized by a
temperature of 4~K and optical depth $4\times 10^{-2}$ at 30 cm$^{-1}$.
We have compared their observations 
to the FIRAS spectrum at the same location. The excess
submillimeter emission from the MIT/Goddard spectrum is 
larger than that seen by FIRAS, but the results are consistent to
within the uncertainties due to different beam sizes and calibration.
Similarly, combining broad-band observations at 6, 9, and 12 cm$^{-1}$ 
(1700, 1100, 830 $\mu$m) by a Santa Barbara/Berkeley balloon experiment
(Meinhold \etal 1993) with IRAS 100$\mu$m observations, 
the best-fitting single-component model for the dust emission
required $\alpha=1.4$ for lines of sight
near the star $\mu$ Peg ($b=-30.7^\circ$). 
A low value of $\alpha$ indicates excess emission with respect to
the $\nu^2$ emissivity expected at low frequencies,
and therefore the Santa Barbara/Berkeley results provide more 
observational evidence for the very cold component in the spectrum
of interstellar dust at high galactic latitude.

\section{Discussion}

\subsection{Warm Dust (16--23 K)}

The temperature of the warm component of the FIRAS spectra agrees with
predictions for dust with a size distribution consistent with 
interstellar extinction (Mathis, Rumpl, \& Nordsieck 1977) and
heated by the interstellar radiation field
(Mathis, Mezger, \& Panagia 1983; Draine \& Anderson 1985). They are
also similar to those derived from broadband 60--100~$\mu$m 
observations by IRAS (Sodroski \etal 1987; Bloeman, Deul, \&
Thaddeus 1990). However, the temperatures inferred from the
60--100~$\mu$m ratios do not show the longitude variation expected
for dust heated by a radiation field that increases toward the
galactic center. This mystery, which has posed problems for models of the
physical properties of the
dust grains responsible for the emission (D\'esert, Boulanger, \&
Puget 1990), is resolved now that all-sky surveys
at longer wavelengths are available. The DIRBE broadband 140/240~$\mu$m 
temperature (Sodroski \etal 1994) shows the same longitude 
variation evident in Figure~\ref{ttaufig}. The FIRAS observations
demonstrate that the spectrum of interstellar dust
emission in the 100--300~$\mu$m range is due to emission from grains
with nearly a single average temperature. The excess mid-infrared emission,
which dominates shortward of 40~$\mu$m and contributes strongly
at 60~$\mu$m, is evidently due to particles undergoing temperature
fluctuations (Draine \& Anderson 1985; D\'esert \etal 1990).

The fact that the interstellar dust spectrum between 100--300~$\mu$m
can be fitted by a single modified blackbody suggests that it is produced
by large grains in equilibrium with the interstellar radiation
field. Thus observations longward of 100~$\mu$m can be reliably used 
to infer the optical depth of interstellar dust. The narrow temperature
range also suggests that the emission is not produced by multiple
grain types with widely different temperatures. For example, a model
with bare graphite and silicate grains ({\it e.g.} Draine \& Lee 1984)
predicts two distinct grain temperatures. For particles with equal
radii of 0.1 $\mu$m, graphite and silicate grains are predicted to 
have temperatures of 18.8~K and 15.4~K, respectively. Taking into
account the predicted abundance and absorption efficiency,
the graphite grains produce about 60--75 \% of the far-infrared emission.
A two-composition model with the above mixture is a significantly
worse fit to the far-infrared spectrum observed by FIRAS
than a single component with free temperature. Further work is needed
to determine the appropriate temperature distribution of the
grains. At this stage we can state that the FIRAS observations
require that the range of temperatures that contribute to the far-infrared
emission is very narrow, and does not allow distinct compositional 
components with comparable abundances.

\subsection{Very Cold Component (4--7 K)}

The emission that we have characterized as a modified blackbody is a
new component of the interstellar spectrum. It would have been difficult
to detect without the continuous spectral coverage of FIRAS,
and it only becomes important at frequencies below 15 cm$^{-1}$;
see Figure~\ref{figcomp}.
In the first paper presenting FIRAS observations of the Galaxy,
Wright \etal found that the very cold emission
was correlated with a galactic angular template based on the total 
intensity in the FIRAS spectra (over the cosmic background radiation).
The very cold component was also found in the FIRAS data by Barnes (1994),
who used a principal component analysis to separate the FIRAS data
into `eigenspectra' and `eigenmaps'.
We have now determined the angular distribution of the very cold component
by modelling coadded spectra regions at high and low latitude.
The salient results for constraining the nature of the very cold component 
are the following:
(1) The temperature of the very cold component varies little with galactic
longitude, and is not correlated with the temperature of the warm dust.
(2) The very cold component is ubiquitous, and its optical depth is correlated
(though not directly proportional to) the optical depth of the warm dust
emission. We consider now some hypotheses for the nature of the 
very cold component.

\subsubsection{Dust shielded from the interstellar radiation field?}

A straightforward hypothesis for the origin of the
very cold component is dust heated
by an attenuated radiation field. For example, dust inside
molecular clouds with extinction $A_V > 1$ is cooler because
it is protected from ultraviolet photons.
In order to explain the
low temperature of the very cold component, the heating rate must
be attenuated by a factor $(T_1/T_2)^6 \sim 10^3$. 
Calculations 
of the penetration of the interstellar radiation field into 
molecular clouds indicate that 
the heating rate is attenuated by a factor of $10^{3}$
at the center of a cloud with extinction $A_V\sim 20$
for silicate grains and $A_V\sim 50$ for graphite grains
(Mathis \etal 1983, hereafter MMP). 
Similar results are obtained by somewhat more detailed models
that take into account the self-heating of the cloud by 
infrared emission from the outer layers (Bernard \etal 1992).
The corresponding column density of
gas is $N_H\sim 4-10\times 10^{22}$~cm$^{-2}$ (Bohlin, Savage,
\& Drake 1978).
By the standards of the diffuse interstellar medium,
this is a large column density, so clouds containing 
shielded dust should be well known.

We note that a much lower estimate for the column density needed to shield
dust to temperatures corresponding to the very cold component
was obtained by Barnes (1994), who concluded that the very cold 
component could be explained by dust shielded by the outer layers
of molecular clouds with $A_V=3$~mag. 
Barnes scaled the intensity of starlight in clouds
by the exponential factor, $e^{-\tau_{UV}}$,
where $\tau_{UV}$ is the far-ultraviolet optical depth.
This estimate neglects the ability of visible photons, which
contain more than half of the energy density of the interstellar 
radiation field,
to penetrate much deeper than UV photons (cf. Mathis \etal 1981). 
Further, scattering allows even ultraviolet radiation to penetrate 
much deeper than the exponential factor predicts; for example, 
starlight in the center of a cloud with total extinction $A_{UV}=20$ 
is predicted to be $\sim 50$ times brighter when scattering 
(with $\omega=g=0.5$) is included (Flannery, Roberge, \& Rybicki 1980).
Very large column densities are needed to shield dust from
the interstellar radiation field.

The very cold component is observed to be ubiquitous; in particular,
it is present in the high-latitude spectra. In order for shielded dust
to explain the very cold component, there must be clouds with high 
extinction distributed throughout the sky. This possibility can be
excluded, and we can show that their filling factor must be small
on angular scales down to $5^\prime$.
Because we can see external galaxies, it is clear that dark clouds 
cannot fill much of the sky on scales larger of a few degrees.
The Shane \& Wirtanen (1967) galaxy counts within
1 deg$^2$ bins are relatively smooth, and are
anticorrelated with the 21-cm line brightness (Heiles 1976).
The bin-to-bin dispersion in the galaxy counts ($\sim 25$\% at
$|b|>20^\circ$) is due primarily to the `intrinsic mottling in 
the surface density of the distribution of galaxies' 
(Burstein \& Heiles 1978). 
Thus on very large scales, there is no evidence for optically thick
clouds that could shield dust from the interstellar radiation field.

On intermediate scales, regions containing shielded dust should be
observable as dark clouds against background starlight.
The Palomar Observatory Sky Survey has been surveyed for all regions 
of noticeable extinction (Lynds 1962; Magnani, Blitz, \& 
Mundy 1985), and the regions of high-latitude extinction were
found to be molecular clouds traced by CO rotational lines. 
However, the total extinctions of these clouds are generally small 
($\langle A_V\rangle \simeq 1$ for individual clouds;
 de Vries \& Magnani 1986). 
Further, the covering fraction of the 
CO-emitting portion of high-latitude molecular clouds is small
($\sim 5\times 10^{-3}$; Magnani, Lada, \& Blitz 1986; but
see also Heithausen \etal 1993)
so that their column density averaged over large regions is
only of order $\langle N_H\rangle \sim 10^{19}$~cm$^{-2}$.
We may relate our derived optical depths at
$\nu_0=30$~cm$^{-1}$ ($333\mu$m) to column densities using
$\tau_0=1.3\times 10^{-25} N_H({\rm cm}^{-2})$
(based on 250$\mu$m observations; Hildebrand 1983). 
The observed optical depth of the cold component at $|b|>30^\circ$ 
corresponds to a column density of 
$\langle N_H\rangle \sim 5\times 10^{20}$~cm$^{-2}$.
This is some two orders of magnitude greater than the amount of 
molecular gas at high latitudes.

On angular scales down to the $5^\prime$ resolution of the 
{\it Infrared Astronomical Satellite} 100$\mu$m sky survey,
regions of shielded dust should be bright infrared emitters.
Shielded regions must reemit all of the energy absorbed from 
the interstellar radiation field; for almost any solid material, 
this emission occurs at mid- to far-infrared wavelengths. 
The surface brightness of the optically thick regions, 
integrated over infrared wavelengths,
must equal the surface brightness of the interstellar radiation field
(cf. Keene 1981; MMP).
The power in the very cold emission is observed to be
negligible, so the energy would be emitted by the warm
component. The range of temperatures observed for the warm
component is not large, so the surface 
brightness is traced (within a factor of 2) by the 100$\mu$m emission.
The surface brightness
of the ISRF is $4\pi J=2.17\times 10^{-2}$ erg cm$^{-2}$ s$^{-1}$
(MMP), which corresponds to a specific surface brightness at 100$\mu$m
for $T=18$~K and $\alpha=2$ of $I_\nu(100\mu{\rm m})=43$ MJy sr$^{-1}$.
It is straightforward to see all regions with 
$I_\nu(100\mu{\rm m})>1$~MJy~sr$^{-1}$ on the IRAS sky maps.
The bright IRAS-detected clouds correspond to known dark clouds and
high-latitude molecular clouds, with some exceptions 
(cf. Blitz, Bazell, \& D\'esert 1990). 
As discussed above, the amount of material in high-latitude
molecular clouds is far too low to explain the observed brightness
of the very cold component of the FIRAS spectra.

On small angular scales, the shielded regions would be visible as
extinction on high-resolution optical images and would be 
point sources to IRAS. Optical images of high-latitude clouds reveal
small-scale structure that does not always correlate with the infrared
emission (Paley \etal 1991, Guhathakurta 1994). The fact that these regions are seen
as {\it bright} on optical images suggests that the light is a combination
of scattered starlight and luminescence. The lack of a 100$\mu$m 
counterpart precludes these regions being optically thick to the
ISRF, so we agree with previous suggestions that the anomalous optical 
brightness is due either to an admixture of large grains or to radiative
transfer ({\it e.g.} shadowing) effects.
The existence of small, shielded regions can be ruled out on thermodynamical
grounds. Such small, shielded regions would have substantial overpressure 
and would rapidly evaporate without a massive central object to hold
them together: the pressure of shielded regions smaller than $5^\prime$
within 100 pc is at least two orders of magnitude larger than the
pressure in the diffuse ISM (cf. Reach, Heiles, \& Koo 1993).


\subsubsection{Very small grains?}

Very small grains ($< 0.02\mu$m) undergo temperature fluctuations when 
exposed to an ultraviolet radiation field, because the mean interval 
between successive ultraviolet photons is longer than the cooling time.
Immediately after an ultraviolet photon is absorbed, very small grains cool
by radiating at frequencies far higher than the larger grains.
In the time interval between ultraviolet
photon impacts, the temperature of the particle is very low, but at least 
the cosmic microwave background temperature (2.7 K). 
Nonequilibrium emission has been invoked to
explain the mid-infrared emission of interstellar dust (Draine \& Anderson
1985; Weiland \etal 1986; D\'esert \etal 1990). 
For example, a 0.001~$\mu$m silicate grain is predicted to reach maximum
temperatures $\sim 200$~K, but to spend the bulk of the time at
temperatures $< 10$~K (Draine \& Anderson 1985). The emission from
very small grains was modeled by D\'esert \etal (1990) from
the near-infrared to 1~mm wavelength. In their model some submillimeter
emission is predicted from the small grains, and it accounts for 12\% of the 
emission at 800~$\mu$m. 
Based on our two-component fits to the spectra
of the outer galactic plane ($270^\circ > l > 90^\circ$), 
the very cold component contains $1/3$ of the total emission 
at 800~$\mu$m. 
If the models have underpredicted the 800~$\mu$m emission of very small
grains by a factor of $\sim 3$, for example because of our poor understanding
of their low-temperature heat capacity, then very small grains could
produce the cold component observed in the FIRAS spectra. 
However, stochastically heated particles producing the 800~$\mu$m
emission will produce excess emission at wavelengths shorter than 60~$\mu$m,
and will produce mid-infrared colors inconsistent with IRAS and DIRBE
observations.
If there were a heating mechanism capable of maintaining very small grains 
at a minimum temperature $\sim 6$~K, where they spend the bulk of their 
time, then they would be able to explain the very cold component;
such a mechanism is not presently known.

\subsubsection{Grains with unusual optical properties?}

A population of cold dust grains, either well mixed with the warm grains
or in widely-distributed clouds, could explain the very cold component.
Balancing heating and cooling, the grain temperature scales
as $T^6-T_{CBR}^6 \propto (Q_{\rm vis}/Q_{\rm FIR}) J_{\rm ISRF}$, where 
$Q_{\rm vis}$ is the absorption efficiency averaged over the 
interstellar radiation field, $Q_{\rm FIR}$ is the absorption
efficiency averaged over the grain's emitted spectrum, 
$J_{\rm ISRF}$ is the strength of the interstellar radiation
field, and we have assumed $\alpha=2$. 
Particles with  $Q_{\rm vis}$/$Q_{\rm FIR}$ a
factor of $10^{-3}$ lower than that of the grains responsible for
the warm ($T\simeq 18$~K) dust emission
could explain the very cold component. 

Fractal grains have enhanced efficiency for submillimeter emission,
and therefore they can cool to very low temperatures (Wright 1993).
A population of fractals may exist such that, when averaged over
their size and shape distributions, the peak emission has characteristic
temperatures corresponding to the observed cold component. 
Because of their high efficiency at emitting in the far-infrared,
even relatively large fractal grains undergo temperature fluctuations
(Bazell \& Dwek 1990, Wright 1993).
A population of fractal grains that spend the bulk of the time
at very low temperature (not much above 2.7 K) and pulse to
5--10 K after absorbing a photon from the radiation field could
produce the observed cold component of the FIRAS spectra.
It is very important that the absorption cross-section per unit mass
of fractal grains is substantially higher than for compact spherical
grains. Thus the mass of fractal dust required to produce the cold 
component is not large. In order to explain the correlation of the
warm and cold components, the fractal grains would have to be spatially
correlated with the grains responsible for the warm component.

\subsubsection{Very large particles?}

Particles sufficiently large that they can cool via photons with
wavelengths much smaller than their radii reach an equilibrium
temperature in the interstellar radiation field such that
$T = (J/\sigma_{SB} + T_{CBR}^4)^{1/4} \simeq 2.9$ K,
where $J$ is the angular-averaged interstellar radiation field
and $\sigma_{SB}$ is the Stefan-Boltzmann constant.
Smaller particles emit with decreased efficiency in the far-infrared,
and thus are warmer. Grains of astronomical silicate and graphite
(Draine \& Lee 1984) equilibrate at 6 K if they have radii of
order 100~$\mu$m.
Similar particles ({\it e.g.} 30$\mu$m amorphous C spheres)
have been proposed before,
in order to explain excess submillimeter absorption in H~II regions
(Rowan-Robinson 1992).
A problem with the hypothesis that large grains produce the very cold
component is that the temperature of the cold component does not
change much in the inner Galaxy, despite a substantially higher
heating rate there.

The mass of large grains required to explain the observed submillimeter
emission is not large. At a wavelength of 800~$\mu$m, the absorption
efficiency of 100~$\mu$m spheres is 0.2 for graphite or 0.4 for silicate
(Draine \& Lee 1984). The mass of dust required to explain the very cold
component is about 0.1\% of the mass of interstellar gas. 
If the large grains contain C or (MgSiFe)O$_4$ groups, their abundance must 
be less than $\sim 0.0073$ relative to H by mass (based on the abundances in
Anders \& Grevesse 1989); this upper limit is a factor of 7 times larger
than the inferred mass of large grains.
Based on the strength of the 10$\mu$m bands seen in absorption
toward bright infrared sources, nearly all of the cosmic Si
abundance is locked in grains smaller than 1$\mu$m;
similarly, infrared absorption bands and the 2200 \AA~ bump require 
$\sim 60$\% of C to be locked in small grains (Tielens \& Allamandola 1987).
This still leaves enough room for the large grains to be composed of 
either silicates or carbonaceous material.

Size distributions of the form $dn/da\propto a^{-m}$, 
up to a maximum size $a_{max}$, have been found to match the wavelength
dependence of the extinction of starlight, where
$m=3.5$ with $a_{max}=0.25$~$\mu$m (Mathis, Rumpl, \& Nordsieck 1977; 
Draine \& Anderson 1985).
If this distribution were to extend from $a_{max}$ to 100~$\mu$m,
there would be 50 times more particles than are needed to explain the
very cold component of the far-infrared spectrum; 
therefore the slope of the size distribution from $a_{max}$ 
to 100~$\mu$m would have to steepen to $m\sim 4.2$.
We note that there are other limits to the 
number of very large particles. For example, inversion of the
wavelength-dependence of polarization into a size distribution
shows that the size distribution must steepen at sizes larger
than $\sim 0.7$~$\mu$m (Kim \& Martin 1994).
Any model for the large particles would have to match the
extinction and polarization curves, while remaining consistent with 
cosmic abundances.

\subsubsection{An emissivity enhancement?}

The cold component we observe may not be cold dust at all, but 
rather a spectral feature in the long-wavelength emissivity of the 
particles that produce the bulk of the far-infrared emission (`warm dust').
Such an emissivity feature could arise from resonances in the grains,
impurities, or molecules in the mantles.
The fact that the mixing ratio of the cold and warm components is
nearly the same at every position motivates this hypothesis. 
In order to test this hypothesis, we divided the observed spectra
by the warm component of the spectral models with $\nu^2$ emissivity
at low frequency. This ratio, the emissivity enhancement,
should reveal the shape of the spectral feature.
The observed spectrum deviates significantly
from the warm component at frequencies below 25 cm$^{-1}$, and the
excess at 12 cm$^{-1}$ is about 30\%.
The enhancement increases toward lower frequencies as $\nu^{-0.35}$,
at least to 10 cm$^{-1}$.

If the excess submillimeter emissivity were due to low-frequency
resonances in the grains, then it would be expected to eventually
disappear at frequencies below the lowest resonant mode.
As a check, we used the DMR observations to check that 
the two-component modified blackbody spectrum does not 
exceed the observed sky brightness at frequencies below those
to which FIRAS was sensitive.
After subtraction of the dipole component of the cosmic
microwave background, 
the correlation slopes of surface brightness 
with csc$|b|$ (for $|b|>15^\circ$) were determined
in the three DMR wavebands; the results are shown 
in Table~\ref{csctab}.
The FIRAS data in each frequency bin treated similarly, 
yielding the spectrum for a cosecant Galaxy model. 
This FIRAS spectrum was fitted by two modified 
blackbodies of the form in Equation~\ref{modbb}, 
with emissivity index $\alpha=2$. 
The warm component
has a temperature $T_1=17.72\pm 0.14$~K and optical depth 
$\tau_1=(1.74\pm 0.06)\times 10^{-5}$, 
and the very cold component has a temperature
$T_2=6.75\pm 0.23$~K and optical depth $\tau_2=(1.23\pm 0.13)\times 10^{-4}$.
The extrapolation of a one-component fit and a two-component 
FIRAS fit down to the DMR frequencies is listed in Table~\ref{csctab}.
Neither dust model exceeds the Galactic emission observed by DMR. 
The DMR observations cannot strongly constrain the low-frequency
dust spectrum, because synchrotron emission, free-free emission, and 
cosmic microwave background anisotropies all contribute to the signal. 
But the fact that the two-component model does not over-predict
the low-frequency emission means that it is possible that the 
excess dust emissivity extends well into the microwave band.
More sensitive microwave observations of the diffuse galactic emission,
and a more detailed study combining the radio, microwave, far-infrared,
and optical (cf. Reynolds 1992) constraints on Galactic and
cosmic sky brightnesses, are
needed to determine the diffuse microwave emission of
interstellar dust.

\subsection{Evidence for other components}

The spectra of the inner Galaxy are poorly
fit even by two modified blackbodies. There is a residual emission
that peaks around 45~cm$^{-1}$ (220~$\mu$m), suggesting a
temperature around 14~K.
This emission is present in spectra within $40^\circ$ of the 
galactic center, so it emanates predominantly from within 
5.5~kpc of the galactic center. 
Comparing the variation of the brightness of this component to that of the
CO(1--0) line from the Goddard-Columbia survey
(Dame \etal 1987) reveals a rough correspondence,
suggesting that this residual emission may be due to dust associated
with molecular clouds.
The temperature of this component is similar to that obtained from
far-infrared spectra of two Bok globules with no local
heating sources (Keene 1981). Thus dust shielded from the interstellar
radiation field {\it is} present in the FIRAS spectra after all, at a
temperature intermediate between that of the `warm' dust and that
of the `cold' dust. The heating rate of the intermediate-temperature
dust is decreased from that of the warm dust by a factor $\sim 5$,
suggesting extinctions $A_V\sim 2$ between the intermediate-temperature
dust and the interstellar radiation field. 
By fitting a three-component spectral model to the data, we found the
typical optical depth of the intermediate-temperature component
is $\sim 1\times 10^{-4}$ at 30~cm$^{-1}$, corresponding to 
a column density $\sim 1\times 10^{21}$~cm$^{-2}$.
The column density (inferred from the optical depth) and the
extinction (inferred from the temperature) are in good agreement
with observations of $N({\rm H})+2N({\rm H}_2)$ and $A_V$ for
interstellar gas and dust (Bohlin, Savage, \& Drake 1978).
This component is very weak, contributing only $\sim 2$\% of
the emission at 45 cm$^{-1}$ and $\sim 60$ times less power
thatn the warm component.

\subsection{Comparison to other galaxies}

Submillimeter and far-infrared observations of other galaxies 
using ground-based and airborne telescopes are challenging with 
current technology,
and continuous spectral coverage of the dust emission comparable to
that obtained by FIRAS for our Galaxy has never been achieved.
But a comparison to observations of external galaxies is important
for two reasons. First, the entire far-infrared emission of external 
galaxies can be measured unambiguously---without confusion from
other emission sources such as zodiacal light and background radiation.
And second, observations of other galaxies allow us to 
determine whether the far-infrared spectral shape is a generic property 
of cosmic dust or is peculiar to individual galaxies. 
Most galaxies observed in the submillimeter range
are significantly warmer than our Galaxy.
Eales, Wynn-Williams, \& Duncan (1989) observed a sample of 10 galaxies
(with large 100$\mu$m fluxes) at 350$\mu$m, with some observations
at 450, 800, \& 1100 $\mu$m as well. Their beamsize corresponded to
3 kpc for the nearest and 30 kpc for the most distant galaxies observed.
The FIRAS spectra of the galactic center and 
inner Galaxy are substantially cooler than the average spectrum of 
Eales \etal: 
the submillimeter fluxes are relatively higher and the 60$\mu$m
flux is lower. In another survey, Clements, Andreani, \& Chase (1993) 
reported observations of 5 galaxies at 450, 800, \& 1100 $\mu$m,
with beamsizes ranging from 1.4--12 kpc.
For the 5 galaxies, the temperature of a single-component fit 
ranged from 28--35~K, compared to the range 16--23~K for spectra
of our Galaxy. 
Recently, Franceschini \& Andreani (1995) reported on a 1.25 mm survey
of a complete sample of southern galaxies, with a selection 
criterion based on the \IRAS\ 60$\mu$m flux. Their average spectrum
is also `warmer' than that of the Milky Way, and they found no
evidence for substantial emission from cold dust.
Evidently the galaxies selected in these surveys  
are warmer than our Galaxy because they were selected
based on their brightness at 60--100~$\mu$m:
Using the \IRAS\ surface brightness to select
galaxies apparently selects the warmest galaxies.
Based on the observed temperatures, it is likely
that the average radiation field within these galaxies 
is nearly an order of magnitude stronger than that in
our own Galaxy.

One other study has shown that the far-infrared spectrum
of the Milky Way is not atypical.
Stark \etal (1989) observed three spirals in the Virgo cluster
at 160$\mu$m and 360$\mu$m with a resolution corresponding to 6.5~kpc. 
When we combine the
FIRAS spectra within $30^\circ$ of the galactic center, a region
that should correspond to what Stark \etal would have observed if 
our Galaxy were at the distance of Virgo, we find that the spectrum
of our Galaxy is entirely consistent with that of the Virgo spirals.

\def\skipme{
Comparison of a $\nu^2$ emissivity to a $\nu^1$
emissivity clearly favored the former. This is in agreement with our
result that $\nu^2$ is a good approximation in the far-infrared, but
it indicates that the cold component may be missing.
}

\subsection{ Implications for Cold Gas as Dark Matter }

The lack of substantial power emitted by dust with temperatures in the 4--15~K
range places strong constraints on the amount of very cold material
in the interstellar medium. The discovery of CO absorption lines from 
clouds in the outer galaxy led to the suggestion that
very cold molecular clouds, which would be very difficult to detect 
by CO emission, may be an important, massive component of the Galaxy
(Lequeux, Allen, \& Guilloteau 1993).
The power of all starlight incident on such cold clouds would be 
reemitted within the FIRAS bandpass.
In fact, there is little power emitted besides that contained in 
the warm component;
the fraction of the total power contained in the cold
component is shown in Figure~\ref{coldpow}. 
Because the cold
component is detected in every direction, it cannot be attributed to
outer galaxy molecular clouds; therefore the power shown in 
Figure~\ref{coldpow} is an upper limit. 
Further, the cold component contains less than 0.05\% of the power 
in the local interstellar radiation field, so that it cannot be
due to optically thick clouds if it is produced locally.

Any model for the Galactic interstellar medium including a substantial
increase in the abundance of dark clouds must comply with the constraint
that very little power is emitted by cold dust.
This constraint has been neglected in some recent work
({\it e.g.} Lequeux \etal 1993; Pfenniger, Combes, \& Martinet 1994a).
One possibility is that such clouds have a very low dust abundance.
This would be consistent with the low gas temperature, because the 
photoelectric heating by dust grains would be absent. 
However, there is evidence for star formation in the far outer galaxy
(De Geus \etal 1993), so there should be dust formation in stellar outflows and
supernovae. Another possibility is that the radiation field is a factor
of $\sim 10^3$ lower in the region of the cold clouds. 
This too is contradicted by the presence of star formation.
Furthermore, power absorbed from the extragalactic radiation field would
still be a factor of $\sim 10$ brighter than our upper limit.

\section{Conclusions}

The continuous spectrum of galactic dust observed by the \COBE/FIRAS can be
characterized as a sum of three emission components: warm dust heated
by the interstellar radiation field, intermediate-temperature dust
heated by the attenuated radiation field in molecular clouds, and
a very cold component. The very cold component 
is correlated with the warm dust, emits very little power, and
its temperature does not increase in the inner Galaxy.
The very cold component is unlikely to be due to dust that is
cold because it is shielded from the interstellar radiation field.
The very cold component may be a spectral feature in the warm
dust constituent material, or may be due to fractal or needle-like grains
or particles substantially larger than those responsible for the
warm component.
The lack of substantial power from cold dust limits the
amount of cold gas in the outer galaxy and makes it unlikely
that such gas could contribute an unseen, massive component of the
Galaxy. 

The existence of enhanced submillimeter emission from 
interstellar dust, regardless of its origin, 
implies an increased contribution to 
the microwave sky brightness by the Galaxy
than previously expected from extrapolations of 100 $\mu$m 
observations ({\it e.g.} by \IRAS). The interstellar dust
emission can be characterized by a two-temperature 
approximation with temperatures and optical depths as
shown in Figure~\ref{mapparfig}.
The implications for searches for {\it fluctuations} in the
microwave background radiation are not yet clear. 
If the cold component is due to enhanced emissivity of the
warm dust constituent material, then its spatial fluctuations
will be comparable to those of an \IRAS\ 100$\mu$m map.
However, if the cold component is due to a separate population
of dust grains in clouds with a substantially different
morphology than previously considered,
the fluctuations in the microwave sky brightness 
due to interstellar dust could be notably enhanced.

\acknowledgements
WTR thanks M. G. Hauser and S. H. Moseley for insightful discussions.

\clearpage
\begin{table}
\vspace{-1truecm}
\caption{Optical depths and temperatures in the galactic plane}\label{planetab} 
\begin{center}
\begin{tabular}{cccccccc}
 & \multicolumn{2}{c}{warm} & \multicolumn{2}{c}{cold} & & &\\
\noalign{\vskip -5truept}
 & \multicolumn{2}{c}{\hrulefill} & \multicolumn{2}{c}{\hrulefill} & & &\\
$l$ ($^\circ$) & $10^5\tau_1$ & $T_1$ (K)& $10^5\tau_2$ & $T_2$ (K) & $\chi^2$/dof & $P(>F_{ac})$ & $P(>F_{bc})$\\
\tableline
$  0$ & $129.1\pm  0.3$ & $25.81\pm 0.02$ & $ 670\pm  72$ & $ 5.4\pm 0.1$ & $ 7.37$ & $>.999$ & $>.999$ \\
$ 15$ & $101.0\pm  0.4$ & $25.02\pm 0.03$ & $ 480\pm  29$ & $ 6.8\pm 0.1$ & $ 2.15$ & $>.999$ & $>.999$ \\
$ 30$ & $ 96.0\pm  0.5$ & $24.69\pm 0.04$ & $ 330\pm  12$ & $ 8.1\pm 0.1$ & $ 3.26$ & $>.999$ & $>.999$ \\
$ 45$ & $ 65.7\pm  0.6$ & $21.88\pm 0.06$ & $ 330\pm  26$ & $ 6.9\pm 0.2$ & $ 1.10$ & $>.999$ & $>.999$ \\
$ 60$ & $ 50.6\pm  0.6$ & $20.59\pm 0.06$ & $ 210\pm  13$ & $ 7.4\pm 0.1$ & $ 1.04$ & $>.999$ & $>.999$ \\
$ 75$ & $ 40.7\pm  1.6$ & $22.18\pm 0.23$ & $ 220\pm  25$ & $ 8.1\pm 0.3$ & $ 0.95$ & $>.999$ & $0.926$ \\
$ 90$ & $ 44.0\pm  1.3$ & $19.86\pm 0.14$ & $ 240\pm  19$ & $ 7.7\pm 0.2$ & $ 1.33$ & $>.999$ & $0.987$ \\
$105$ & $ 29.9\pm  0.6$ & $20.57\pm 0.10$ & $ 160\pm   9$ & $ 7.9\pm 0.2$ & $ 1.10$ & $>.999$ & $>.999$ \\
$120$ & $ 31.9\pm  0.6$ & $19.63\pm 0.09$ & $ 190\pm  11$ & $ 7.5\pm 0.1$ & $ 1.07$ & $>.999$ & $>.999$ \\
$135$ & $ 23.5\pm  0.6$ & $21.33\pm 0.12$ & $ 140\pm   7$ & $ 8.3\pm 0.2$ & $ 1.65$ & $>.999$ & $0.945$ \\
$150$ & $ 31.4\pm  1.0$ & $19.12\pm 0.15$ & $ 180\pm  16$ & $ 7.5\pm 0.2$ & $ 1.12$ & $>.999$ & $0.973$ \\
$165$ & $ 16.8\pm  1.0$ & $18.87\pm 0.27$ & $ 100\pm  13$ & $ 7.7\pm 0.4$ & $ 1.03$ & $>.999$ & $0.760$ \\
$180$ & $ 23.8\pm  1.1$ & $18.72\pm 0.21$ & $ 160\pm  24$ & $ 6.9\pm 0.3$ & $ 0.90$ & $>.999$ & $0.917$ \\
$195$ & $ 28.7\pm  3.1$ & $18.58\pm 0.47$ & $ 160\pm  43$ & $ 7.4\pm 0.7$ & $ 1.01$ & $0.928$ & $0.593$ \\
$210$ & $ 23.5\pm  1.3$ & $18.94\pm 0.25$ & $ 130\pm  17$ & $ 7.7\pm 0.3$ & $ 1.04$ & $>.999$ & $0.640$ \\
$225$ & $ 17.0\pm  1.5$ & $20.23\pm 0.40$ & $  80\pm   8$ & $ 9.0\pm 0.4$ & $ 1.12$ & $>.999$ & $0.432$ \\
$240$ & $ 16.3\pm  1.4$ & $19.12\pm 0.38$ & $ 110\pm  15$ & $ 7.9\pm 0.4$ & $ 0.87$ & $>.999$ & $0.608$ \\
$255$ & $ 28.9\pm  1.4$ & $18.85\pm 0.21$ & $ 140\pm  20$ & $ 7.6\pm 0.4$ & $ 0.95$ & $>.999$ & $0.945$ \\
$270$ & $ 39.6\pm  0.6$ & $20.91\pm 0.08$ & $ 220\pm   8$ & $ 8.2\pm 0.1$ & $ 1.84$ & $>.999$ & $>.999$ \\
$285$ & $ 38.0\pm  0.4$ & $23.94\pm 0.06$ & $ 250\pm   6$ & $ 8.6\pm 0.1$ & $ 2.67$ & $>.999$ & $>.999$ \\
$300$ & $ 51.5\pm  0.5$ & $22.48\pm 0.06$ & $ 280\pm  14$ & $ 7.6\pm 0.1$ & $ 1.83$ & $>.999$ & $>.999$ \\
$315$ & $ 79.3\pm  0.2$ & $23.10\pm 0.02$ & $ 430\pm  15$ & $ 6.6\pm 0.1$ & $ 2.33$ & $>.999$ & $>.999$ \\
$330$ & $112.0\pm  0.2$ & $24.42\pm 0.01$ & $ 660\pm  32$ & $ 5.8\pm 0.1$ & $ 4.68$ & $>.999$ & $>.999$ \\
\end{tabular}
\end{center}
\end{table}
\clearpage

\begin{table}
\caption{Optical depths and temperatures at high latitude}\label{highlattab} 
\begin{center}
\noindent\hskip-1truecm\begin{tabular}{cccccccc}
 \multicolumn{8}{l}{\it (a) Single-component fit ($\alpha=2$)}\\
region   & $10^5\tau$ & $T$ (K)& $\chi^2$/dof & & & &\\
\multicolumn{8}{l}{\hrulefill}\\
$b>60$ & $ 0.24\pm 0.00$ & $15.31\pm 0.08$ & $ 7.56$ & & & &\\
$  0<l< 90, 30<b<60$ & $ 0.30\pm 0.00$ & $15.99\pm 0.06$ & $ 7.07$ & & & &\\
$  0<l< 90, -30>b>-60$ & $ 0.32\pm 0.01$ & $16.47\pm 0.10$ & $ 3.16$ & & & &\\
$135<l<180, 10<b<30$ & $ 0.95\pm 0.01$ & $16.28\pm 0.05$ & $ 3.58$ & & & &\\
$135<l<180, -10>b>-30$ & $ 1.26\pm 0.01$ & $16.66\pm 0.03$ & $ 6.43$ & & & &\\
$180<l<225, 10<b<30$ & $ 0.46\pm 0.01$ & $15.82\pm 0.10$ & $ 3.06$ & & & &\\
$180<l<225, -10>b>-30$ & $ 1.14\pm 0.01$ & $17.91\pm 0.03$ & $ 6.70$ & & & &\\
[0.15 truein]
%
 \multicolumn{8}{l}{\it (b) Single-component fit ($\alpha$ free)}\\
region   & $10^5\tau$ & $T$ (K) & $\alpha$ & $\chi^2$/dof & $P(>F_{ab})$ & & \\
\multicolumn{8}{l}{\hrulefill}\\
$b>60$ & $ 0.08\pm 0.00$ & $23.2\pm 0.5$ & $0.92\pm 0.04$ & $ 4.49$ & $>.99$ & & \\
$  0<l< 90, 30<b<60$ & $ 0.13\pm 0.00$ & $22.2\pm 0.4$ & $1.12\pm 0.03$ & $ 3.00$ & $>.99$ & & \\
$  0<l< 90, -30>b>-60$ & $ 0.18\pm 0.01$ & $20.7\pm 0.5$ & $1.36\pm 0.06$ & $ 2.37$ & $0.95$ & & \\
$135<l<180, 10<b<30$ & $ 0.60\pm 0.02$ & $19.3\pm 0.2$ & $1.51\pm 0.03$ & $ 1.88$ & $>.99$ & & \\
$135<l<180, -10>b>-30$ & $ 0.88\pm 0.01$ & $19.1\pm 0.1$ & $1.60\pm 0.02$ & $ 2.85$ & $>.99$ & & \\
$180<l<225, 10<b<30$ & $ 0.20\pm 0.01$ & $22.0\pm 0.6$ & $1.12\pm 0.06$ & $ 1.78$ & $>.99$ & & \\
$180<l<225, -10>b>-30$ & $ 0.79\pm 0.01$ & $20.8\pm 0.1$ & $1.58\pm 0.02$ & $ 2.68$ & $>.99$ & & \\
\end{tabular}
\end{center}
\end{table}
\clearpage

\begin{table}
\vspace{1cm}
\begin{center}
{Table~\ref{highlattab}. (continued)\hfil}
\end{center}
\noindent\hskip -2truecm \begin{tabular}{cccccccc}
\multicolumn{8}{l}{\hrulefill}\\
\multicolumn{8}{l}{\it (c) Two-component fit ($\alpha=2$)}\\
region & $10^5\tau_1$ & $T_1$ (K) & $10^5\tau_2$ & $T_2$ (K) & 
         $\chi^2$/dof & $P(>F_{ac})$ & $P(>F_{bc})$ \\
\multicolumn{8}{l}{\hrulefill}\\
$b>60$ & $ 0.13\pm 0.01$ & $17.5\pm 0.3$ & $ 1.2\pm 0.2$ & $ 7.1\pm 0.3$ & $ 4.50$ & $>.99$ & $0.50$  \\
$  0<l< 90, 30<b<60$ & $ 0.19\pm 0.01$ & $17.8\pm 0.2$ & $ 1.2\pm 0.1$ & $ 7.3\pm 0.3$ & $ 3.02$ & $>.99$ & $0.48$  \\
$  0<l< 90, -30>b>-60$ & $ 0.25\pm 0.02$ & $17.5\pm 0.2$ & $ 1.4\pm 0.4$ & $ 6.7\pm 0.6$ & $ 2.31$ & $0.96$ & $0.56$  \\
$135<l<180, 10<b<30$ & $ 0.83\pm 0.02$ & $16.8\pm 0.1$ & $ 4.9\pm 1.0$ & $ 5.7\pm 0.3$ & $ 1.66$ & $>.99$ & $0.76$  \\
$135<l<180, -10>b>-30$ & $ 1.16\pm 0.01$ & $17.0\pm 0.1$ & $ 9.5\pm 1.3$ & $ 5.1\pm 0.2$ & $ 1.47$ & $>.99$ & $>.99$  \\
$180<l<225, 10<b<30$ & $ 0.34\pm 0.02$ & $17.0\pm 0.2$ & $ 3.3\pm 0.8$ & $ 6.1\pm 0.4$ & $ 1.72$ & $>.99$ & $0.58$  \\
$180<l<225, -10>b>-30$ & $ 1.04\pm 0.01$ & $18.3\pm 0.1$ & $ 7.9\pm 1.0$ & $ 5.5\pm 0.2$ & $ 1.70$ & $>.99$ & $>.99$  \\
\end{tabular}
\end{table}
\clearpage

\begin{table}
\vspace{-1truecm}
\caption{Comparison of dust models to DMR observations}\label{csctab} 
\begin{center}
\begin{tabular}{ccccc}
 & & & \multicolumn{2}{c}{FIRAS dust models} \\
\multicolumn{2}{c}{frequency}  & DMR csc$|b|$ slope & one-component & two-component \\
(GHz)&(cm$^{-1}$) & (kJy sr$^{-1}$)  & (kJy sr$^{-1}$) & (kJy sr$^{-1}$)\\
\tableline
 31  &  1.0 & $1.5\pm 0.2$ & 0.01 & 0.04 \\
 53  &  1.8 & $1.3\pm 0.2$ & 0.1  & 0.3 \\
 90  &  3.0 & $2.5\pm 0.8$ & 0.9  & 2.2 \\
\end{tabular}
\end{center}
\end{table}

\clearpage

\begin{figure}
\caption{The FIRAS spectrum of the region $(85^\circ < l < 95^\circ,
25^\circ < b < 35^\circ)$ is shown together with its decomposition
into cosmic background radiation (CMBR), Galactic emission, and
zodiacal light.}
\label{figcgz}
\end{figure}

\begin{figure}
\caption{{\it (a)} Spectrum of interstellar emission in 
the galactic plane toward longitude $l=45^\circ$. Other than
the bright spectral lines are due to C$^+$ and N$^+$, the spectrum
is dominated by emission from warm dust, which peaks around
65 cm$^{-1}$ (150$\mu$m).
{\it (b)} Residuals after a single-component modified blackbody
with $\alpha=2$ was subtracted. The residual intensity at each 
frequency was divided by the uncertainty at that frequency. The
excess emission at 7--20 cm$^{-1}$ (1400--500$\mu$m) is evident.
{\it (c)} Residuals after a single component with the best-fitting
emissivity index was subtracted. The fit has improved with the extra
degree of freedom, but the true shape of the spectrum at low 
frequencies has not been accurately fitted.
{\it (d)} Residuals after a two-component model was subtracted.
The very cold emission is now adquately fit. The spectral lines
of CO(4--3) and CI can be clearly seen. There is evidence for
excess emission in the 40--55 cm$^{-1}$ (250--180$\mu$m) for the
inner galaxy spectra.}
\label{fitfig}
\end{figure}

\begin{figure}
\caption{{\it (a)} Spectrum of interstellar emission in 
the region with $-30>b>-60^\circ$ and $180^\circ>l>90^\circ$.
Only a weak C$^+$ spectral line and the warm dust continuum are evident.
{\it (b)} Residuals after a single-component modified blackbody
with $\alpha=2$ was subtracted. 
The excess emission at 7--20 cm$^{-1}$ (1400--500$\mu$m) is evident.
{\it (c)} Residuals after a single component with the best-fitting
emissivity index ($\alpha=1.4$) was subtracted.
{\it (d)} Residuals after a two-component model was subtracted.
It is not possible to clearly distinguish whether panel (c) or (d) are
better fits to the data.}
\label{fitfigh}
\end{figure}

\begin{figure}
\caption{Longitude profiles of the optical depth and temperature of the
two-component, four-parameter fits to the spectra of the galactic
plane.}
\label{ttaufig}
\end{figure}

\begin{figure}
\caption{Maps showing the variation of the temperature and optical
depth of the `warm' and `cold' components. The maps are in galactic
coordinates, with the galactic center in the middle. The grid
lines delineate the regions used for spectral fitting, and the values
listed within each region are the fitted parameters and their statistical 
uncertainties. Across the galactic plane, the longitude variation from
the previous Figure is repeated for comparison.}
\label{mapparfig}
\end{figure}

\begin{figure}
\caption{Optical depth of the cold component {\it versus} optical
depth of the warm component for all spectra. The two components are
well correlated, which strongly suggests that the cold component is
galactic emission. Because the galactic column density varies strongly
with latitude, the points in the lower-left-hand corner are all at
high galactic latitudes. The cluster of points in the upper-right-hand
corner are the galactic plane spectra; the outer galactic spectra
are at the lower left of this cluster and the galactic center is at
the top right. The line representing $\tau_2/\tau_1=7$ is shown
for comparison.}
\label{tautaufig}
\end{figure}

\begin{figure}
\caption{The FIRAS galactic spectrum of the region $(l=45^\circ,
|b|<3^\circ)$ is shown together with its decomposition
into warm and very cold components.}
\label{figcomp}
\end{figure}

\begin{figure}
\caption{Fraction of the total power contained in the `cold' component,
as a function of galactic longitude.}
\label{coldpow}
\end{figure}

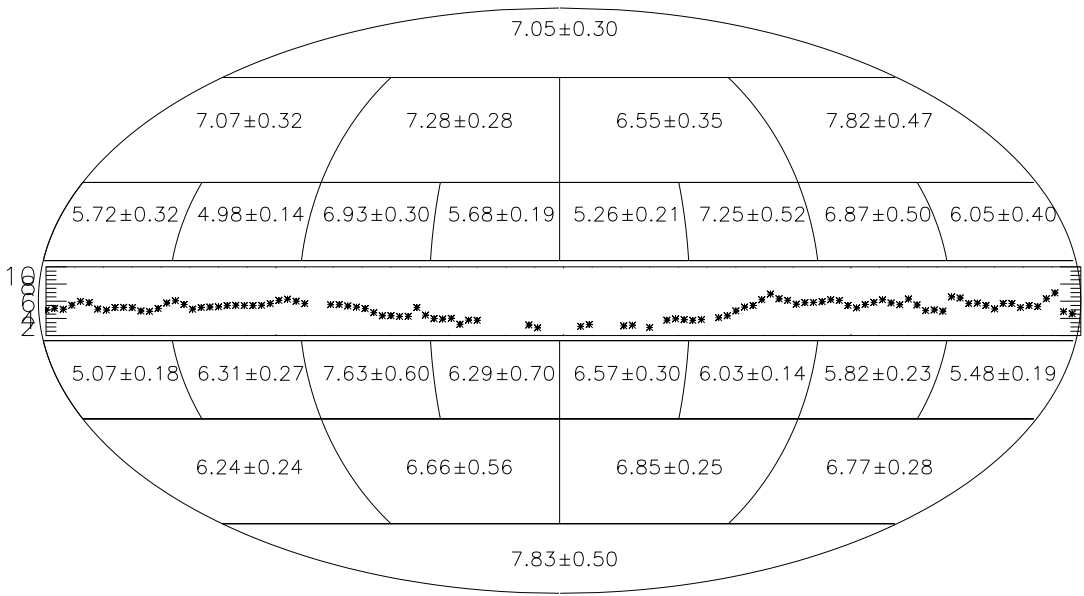
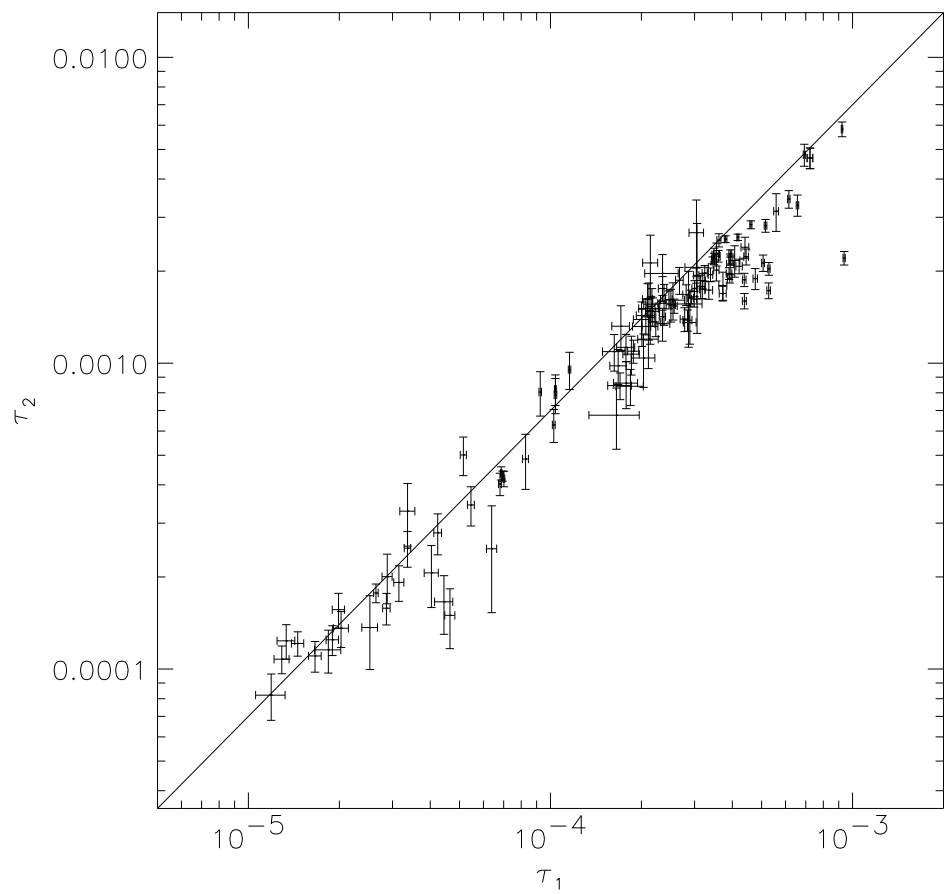
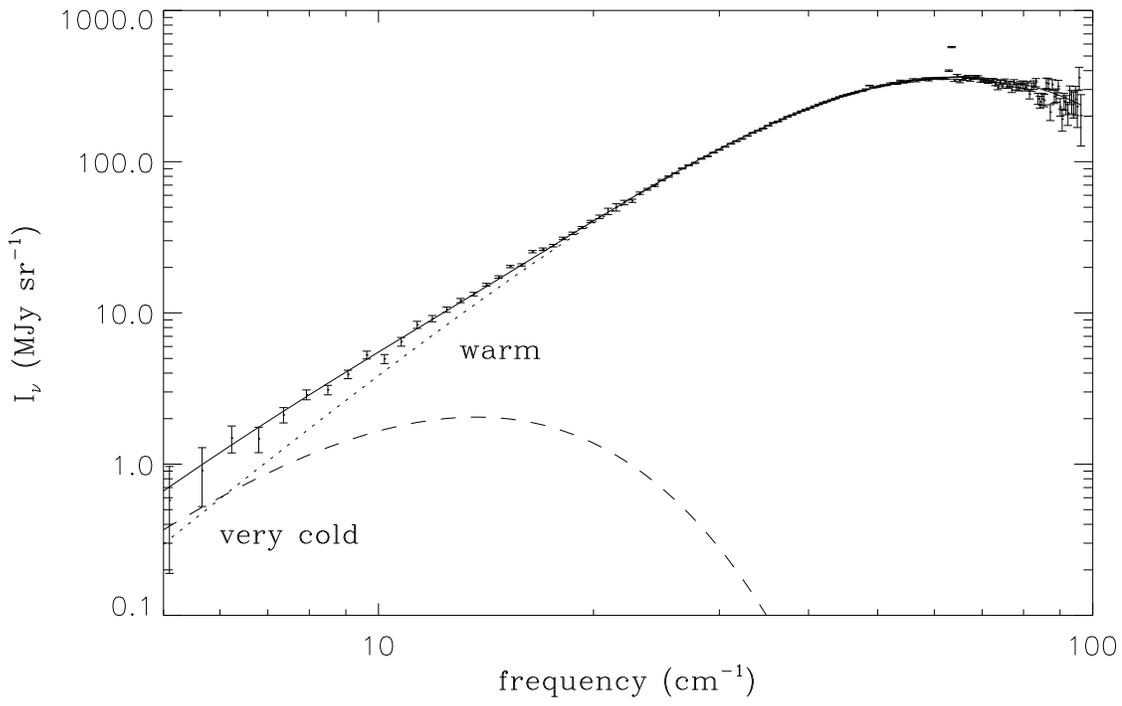
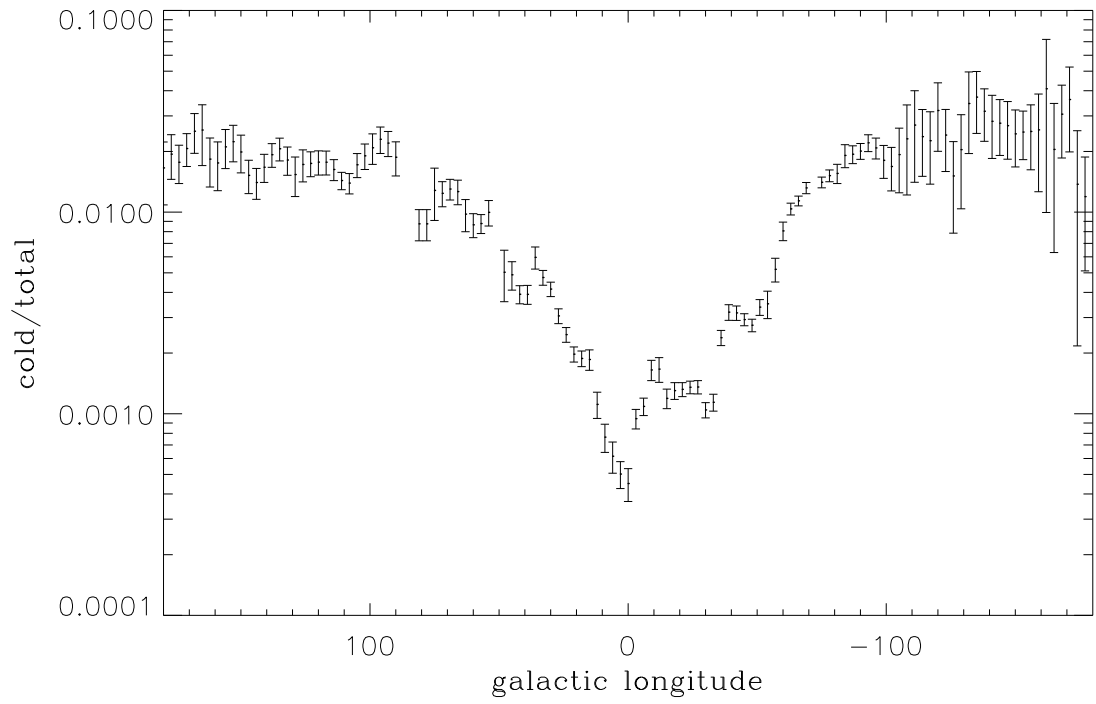
\end{document}